\documentstyle[preprint,tighten,aps,epsf]{revtex}
\begin{document}
\draft
\title{Bosonization and the eikonal
expansion: similarities and differences}
\author{Peter Kopietz}
\address{
Institut f\H{u}r Theoretische Physik der Universit\H{a}t G\H{o}ttingen,\\
Bunsenstr.9, D-37073 G\H{o}ttingen, Germany}
\date{January 23, 1996}
\maketitle
\begin{abstract}
We compare two non-perturbative techniques for calculating
the single-particle Green's function of interacting
Fermi systems with dominant forward scattering:
our recently developed functional integral approach to bosonization
in arbitrary dimensions, and the eikonal expansion.
In both methods the Green's function is first calculated
for a fixed configuration of a background field, and then
averaged with respect to a suitably defined effective action. 
We show that, after linearization of the
energy dispersion at the Fermi surface,
both methods yield for Fermi liquids
exactly the same non-perturbative expression for the
quasi-particle residue.
However, in the case of  non-Fermi liquid  behavior
the low-energy behavior of the Green's function predicted
by the eikonal method can be erroneous.
In particular, for the Tomonaga-Luttinger model the eikonal method neither
reproduces the correct scaling behavior of the
spectral function, nor predicts the correct location
of its singularities.

\end{abstract}
\narrowtext

\section{Introduction}
Recently a number of authors have developed generalizations of the bosonization 
method to arbitrary dimensions\cite{Luther79,Haldane92,%
Houghton93,Kwon95,Castro94,Frohlich95,Kopietz94,Kopietz95,Kopietzorb,Kopietzhab}.
This approach is most suitable in interacting Fermi systems where
the Fourier transform of the 
interaction is dominated by momentum transfers
small compared with the Fermi momentum $k_F$. 
In this case the single-particle Green's function satisfies
a non-trivial asymptotic Ward-identity\cite{Castellani94,Castellani94b}, which opens the
way for a controlled summation of the entire
perturbation series. Bosonization is 
one possible (and perhaps the most efficient) way to explicitly
carry out this summation.
Interactions that exhibit in Fourier space 
a strong maximum or
even a singularity for small momentum transfers
are long-range in real space. The experimentally perhaps
most  relevant interaction of this type is the effective
current-current interaction due to the
coupling between fermions and transverse gauge fields.
Such a coupling appears in models for strongly
correlated electron systems\cite{Lee89}.
Moreover, effective current-current interactions
are also generated by fluctuations of the statistical gauge field
in the Chern-Simons theory for quantum Hall systems\cite{Halperin93}.
In Refs.\cite{Kwon95,Kopietz95rad} the
gauge field problem has been studied via
higher-dimensional bosonization. 
However, this approach has been criticized\cite{Ioffe94,Castellani94b},
because higher-dimensional
bosonization relies (at least in its simplest form)
on the linearization of the energy dispersion
at the Fermi surface, and there exists evidence that in the
gauge field problem the curvature of the Fermi surface
cannot be neglected\cite{Ioffe94,Castellani94b}.

An alternative non-perturbative approach, which does
not rely on the linearization of the energy dispersion,
is the eikonal method.
This approach has been developed in the 60's to obtain non-perturbative
results for the single-particle Green's function
of field theories
where the fermionic degrees of freedom are linearly coupled to
another bosonic quantum field\cite{Svidzinskii57,Barbashov65,Fradkin66,Popov83}. 
More recently, Khveshchenko and Stamp\cite{Khveshchenko93} 
have used this method to
study the above mentioned gauge field problem in two dimensions. 
The strategy is to calculate first the
Green's function
${\cal{G}} ( {\bf{r}} , {\bf{r}}^{\prime} , t , t^{\prime} ; \{ {\bf{A}} \} )$
of the fermions for fixed configuration of the gauge field ${\bf{A}}$,
and then average the result with respect to  the effective
gauge field action $\tilde{ S}_{eff} \{ {\bf{A}} \}$ to obtain the
translationally invariant fermionic Green's function\cite{footnoteG},
 $G ( {\bf{r}} - {\bf{r}}^{\prime} , t - t^{\prime} )
 = \left<
 {\cal{G}} ( {\bf{r}} , {\bf{r}}^{\prime} , t , t^{\prime} ; \{ {\bf{A}} \} )
 \right>_{\tilde{S}_{eff}  }$.
To leading order in the eikonal expansion the
averaging involves a trivial Gaussian integration, and
gives rise to the usual Debye-Waller factor\cite{Khveshchenko93}.
The functional integral approach to 
bosonization\cite{Fogedby76,Kopietz94,Kopietz95,Kopietzorb,Kopietzhab,Kopietz95rad} 
follows precisely the same
strategy: The fermionic two-body interaction is first decoupled via a 
Hubbard-Stratonovich transformation involving a space--  and time-dependent
auxiliary field; then the fermionic Green's function
is calculated for a fixed configuration of the auxiliary field, and finally
the result is averaged with respect to an effective
action $S_{eff}$.
Although the functional bosonization 
approach\cite{Fogedby76,Kopietz94,Kopietz95,Kopietzorb,Kopietzhab,Kopietz95rad}
is formulated in Euclidean (i.e. imaginary) time
while the eikonal expansion is by construction a
Minkowski (i.e. real) time method,
the similarity in both procedures suggests that,
after proper analytic continuation, they should yield 
equivalent results for models with linearized energy dispersion.
In the present paper we shall carefully examine 
this point, 
and show that in general this is not the case.


\section{Functional bosonization}
\label{sec:bos}

In order to compare bosonization with the eikonal expansion, we shall
in this section briefly describe our
functional bosonization approach, focusing on the calculation of the
single-particle Green's function. For a more
detailed description, see Refs.\cite{Kopietz94,Kopietz95,Kopietzorb,Kopietzhab}.
To calculate Matsubara Green's function of the many-body system, we
subdivide the degrees of
freedom close to the Fermi surface into a finite number of boxes $K^{\alpha}$ 
labelled by an index $\alpha$, 
and decouple the interaction with the help of a 
Hubbard-Stratonovich transformation involving   
auxiliary fields associated with the $K^{\alpha}$.
For our purpose it is sufficient to consider
conventional density-density interactions, which can be
decoupled via scalar fields $\phi^{\alpha}$. 
The generalization to gauge fields is straightforward and can
be found in Refs.\cite{Kopietzhab,Kopietz95rad}.
After the standard transformations\cite{Kopietzorb,Kopietzhab}
the Matsubara Green's function $G ( k ) \equiv G ( {\bf{k}}, i \tilde{\omega}_n )$ can
be {\it{exactly}} written as\cite{footnote1}
 \begin{eqnarray}
 G( k )  = 
 \int {\cal{D}} \{ \phi^{\alpha} \} 
 {\cal{P}} \{ \phi^{\alpha} \} 
 [ \hat{G} ]_{kk}
 \equiv 
 \left< [ \hat{G}]_{  k  k }   \right>_{ S_{eff} }
 \label{eq:avphi}
 \; \; \; .
 \end{eqnarray}
Here $\hat{G}^{-1}$ is an infinite matrix in momentum-- and frequency space, with
matrix elements given by the formal Dyson equation
$ [ \hat{G}^{-1} ]_{k k^{\prime}} =
 [ \hat{G}^{-1}_{0} ]_{k k^{\prime}} - [ \hat{V} ]_{k k^{\prime}}$,
where $\hat{G}_{0}$ is the non-interacting Matsubara Green's function matrix,
$ [ \hat{G}_{0} ]_{ k k^{\prime} }
   =  \delta_{k k^{\prime} } 
[ i \tilde{\omega}_{n} - \epsilon_{\bf{k}} + \mu ]^{-1}$,
and the generalized self-energy matrix $\hat{V}$ is
$ [ \hat{V} ]_{ k k^{\prime} }   =
 \sum_{\alpha} 
   \Theta^{\alpha} ( {\bf{k}} ) V^{\alpha}_{k- k^{\prime}}$, with
$  V^{\alpha}_{q} =
      \frac{i}{\beta} 
  {\phi}^{\alpha}_{q} $.
Here $\beta$ is the inverse temperature, $\mu$ is the chemical
potential, $\epsilon_{\bf{k}}$ is the non-interacting energy dispersion,
and the cutoff function $\Theta^{\alpha} ( {\bf{k}} )$ is unity
if ${\bf{k}}$ belongs to box $K^{\alpha}$, and vanishes otherwise.
The normalized probability distribution
 ${\cal{P}} \{ \phi^{\alpha} \} $ is 
 \begin{equation}
 {\cal{P}} \{ \phi^{\alpha} \} 
 = 
 \frac{  
 e^{ - {S}_{eff} \{ \phi^{\alpha} \} }  }
 {
 \int {\cal{D}} \left\{ \phi^{\alpha} \right\} 
 e^{ - {S}_{eff} \{ \phi^{\alpha} \} }  }
 \; \; \; ,
 \label{eq:probabphidef}
 \end{equation}
where the effective action
for the $\phi^{\alpha}$-field  is of the form
$ {S}_{eff} \{ \phi^{\alpha} \} 
 = {S}_{2} \{ \phi^{\alpha} \} 
 +  {S}_{kin} \{ \phi^{\alpha} \} $,
with
 \begin{eqnarray}
 {S}_{2} \{ {{\phi}}^{\alpha}  \} & = &
 \frac{1}{2} \sum_{q} \sum_{\alpha \alpha^{\prime} }
  [ \underline{\tilde{f}}_{ {{q}} }^{-1} ]^{ \alpha \alpha^{\prime} }
 \phi_{-q}^{\alpha} \phi_{q}^{\alpha^{\prime}}
 \label{eq:S2decoupdef}
 \; \; \; ,
 \\
{S}_{kin} \{ \phi^{\alpha} \}  & = &
 -  Tr \ln [ 1 - \hat{G}_{0} \hat{V} ]  
 \label{eq:Skinphidef}
 \; \; \; .
 \end{eqnarray}
Here $ \underline{\tilde{f}}_{ {{q}} }$ is a matrix in the 
patch indices, with matrix elements 
$ [ \underline{\tilde{f}}_{ {{q}} }]^{\alpha \alpha^{\prime}}
= \frac{\beta}{V} f_q^{\alpha \alpha^{\prime}}$, where
$f_q^{\alpha \alpha^{\prime}}$ are the usual Landau interaction
parameters, and $V$ is the volume of the system.
The probability distribution ${\cal{P}} \{ \phi^{\alpha} \}$ 
can be calculated perturbatively by expanding Eq.\ref{eq:Skinphidef}
in powers of the $\phi^{\alpha}$-field. The validity of this expansion
is controlled by the generalized closed loop theorem, which
is discussed in detail in Refs.\cite{Kopietz95,Kopietzhab}.
At the level of the Gaussian approximation one obtains
 \begin{equation}
   S_{kin} \{ \phi^{\alpha} \}
  \approx 
 \frac{V}{2 \beta } \sum_{q} \sum_{\alpha \alpha^{\prime} }
  {{\Pi}}_{0}^{ \alpha \alpha^{\prime} } ( q )
 \phi_{-q}^{\alpha} \phi_{q}^{\alpha^{\prime}}
 \label{eq:Seff2phigaussres}
 \; \; \; ,
 \end{equation}
where the ``patch'' polarization is given by
 \begin{eqnarray}
  {{\Pi}}_{0}^{ \alpha \alpha^{\prime} } ( q )
  & = & - \frac{1}{2 \beta V} \sum_k
  \left[
  \Theta^{\alpha} ( {\bf{k}} )
  \Theta^{\alpha^{\prime}} ( {\bf{k}} + {\bf{q}} )
  G_0 ( k ) G_{0} ( k +  q )
  \nonumber
  \right.
  \\
  & & 
  \left.
  +
  \Theta^{\alpha^{\prime}} ( {\bf{k}} )
  \Theta^{\alpha} ( {\bf{k}} - {\bf{q}} )
  G_0 ( k ) G_{0} ( k -  q )
  \right]
  \; \; \; .
  \label{eq:patchpol}
  \end{eqnarray}
The  leading correction to
Eq.\ref{eq:Seff2phigaussres} is given in Refs.\cite{Kopietz95,Kopietzhab}.
The calculation of the diagonal elements
$[ \hat{G} ]_{kk}$ in Eq.\ref{eq:avphi} is more difficult.
Choosing the patches larger than the range of the interaction in momentum
space, we may write\cite{Kopietzorb,Kopietzhab}
 $[ \hat{G} ]_{k k} = \sum_{\alpha} 
 \Theta^{\alpha} ( {\bf{k}} ) 
 [ \hat{G}^{ \alpha } ]_{k k }$, where
(after shifting ${\bf{k}} = {\bf{k}}^{\alpha} + {\bf{q}}$,
${\bf{k}}^{\prime} = {\bf{k}}^{\alpha} + {\bf{q}}^{\prime}$)  
the infinite matrix $\hat{G}^{\alpha}$ satisfies\cite{footnote1}
 \begin{equation}
 \sum_{\tilde{q}^{\prime}} \left[ \delta_{\tilde{q} , \tilde{q}^{\prime}  } 
 [ G^{\alpha}_{0} ( \tilde{q} ) ]^{-1}
 - V^{\alpha}_{ \tilde{q} - \tilde{q}^{\prime} } \right] 
 [ \hat{G}^{\alpha} ]_{ \tilde{q}^{\prime} \tilde{q}^{\prime \prime} }
 = \delta_{ \tilde{q} , \tilde{q}^{\prime \prime} }
 \label{eq:Galphadif}
 \; \; \; .
 \end{equation}
Here $[G_{0}^{ \alpha} ( \tilde{q} )]^{-1} = i \tilde{\omega}_{n}
- {\xi}^{\alpha} ({\bf{q}})$,
with $\xi^{\alpha} ( {\bf{q}} ) = \epsilon_{ {\bf{k}}^{\alpha}
+ {\bf{q}}} - \mu$.
Defining
 $V^{\alpha} ( {\bf{r}} , \tau )  =  \sum_{q}
 e^{ i ( {\bf{q}} \cdot {\bf{r}} - {\omega}_{m} \tau )}  V^{\alpha}_{q}$ and
 \begin{equation}
 {\cal{G}}^{\alpha} ( {\bf{r}} , {\bf{r}}^{\prime} , \tau , \tau^{\prime} )
  =  \frac{1}{ \beta V} \sum_{ \tilde{q} \tilde{q}^{\prime}}
 e^{ i ( {\bf{q}} \cdot {\bf{r}} - \tilde{\omega}_{n} \tau )} 
 e^{ - i ( {\bf{q}}^{\prime} \cdot {\bf{r}}^{\prime} - \tilde{\omega}_{n^{\prime}} \tau^{\prime} )} 
 [ \hat{G}^{\alpha} ]_{\tilde{q} \tilde{q}^{\prime} }
 \; \; \; ,
 \label{eq:calGdef}
 \end{equation}
it is easy to see that Eq.\ref{eq:Galphadif} is equivalent with 
 \begin{equation}
  \left[- {\partial}_{ \tau} -    
  {\xi}^{\alpha} ( {\bf{P}}_{\bf{r}} ) - V^{\alpha}  ( {\bf{r}} , \tau )
  \right]
 {\cal{G}}^{\alpha} ( {\bf{r}} , {\bf{r}}^{\prime} , \tau , \tau^{\prime} )
 = \delta ( {\bf{r}} - {\bf{r}}^{\prime} ) \delta^{\ast} ( \tau - \tau^{\prime} )
 \; \; \; ,
 \label{eq:Galphadifrt}
 \end{equation}
where ${\bf{P}}_{\bf{r}} = - i \nabla_{\bf{r}}$ is the momentum operator, and
$ \delta^{\ast} ( \tau - \tau^{\prime} )
 = \frac{1}{\beta} \sum_{n} e^{ - i \tilde{\omega}_{n} ( \tau - \tau^{\prime} ) }$.
Eq.\ref{eq:Galphadifrt} together with the
Kubo-Martin-Schwinger (KMS) boundary conditions\cite{Kubo57},
 \begin{equation}
 {\cal{G}}^{\alpha} ( {\bf{r}} , {\bf{r}}^{\prime} , \tau + \beta , \tau^{\prime} )
  =  
 {\cal{G}}^{\alpha} ( {\bf{r}} , {\bf{r}}^{\prime} , \tau  , \tau^{\prime} + \beta )
 =
  -
 {\cal{G}}^{\alpha} ( {\bf{r}} , {\bf{r}}^{\prime} , \tau , \tau^{\prime} )
 \label{eq:KMS} 
 \; \; \; ,
 \end{equation}
uniquely determine the Green's function.
For {\it{linearized energy dispersion}}, $\xi^{\alpha} ({\bf{q}}) \approx
{\bf{v}}^{\alpha} \cdot {\bf{q}}$, Eq.\ref{eq:Galphadifrt}
can be solved exactly by means of a trivial generalization of
Schwinger's ansatz\cite{Schwinger62}. The result 
is\cite{Kopietz94,Kopietz95,Kopietzorb,Kopietzhab}
 \begin{equation}
 {\cal{G}}^{\alpha} ( {\bf{r}} , {\bf{r}}^{\prime} , \tau , \tau^{\prime} )
  = 
 {{G}}^{\alpha}_{0} ( {\bf{r}} - {\bf{r}}^{\prime} , \tau - \tau^{\prime} )
 e^{ \Phi^{\alpha} ( {\bf{r}} , \tau ) - \Phi^{\alpha} ( {\bf{r}}^{\prime} , \tau^{\prime} ) }
 \label{eq:Ansatz}
 \; \; \;  ,
 \end{equation}
 \begin{eqnarray}
 {{G}}^{\alpha}_{0} ( {\bf{r}}  , \tau  )
 & = &
 \frac{1}{\beta V} \sum_{ \tilde{q}} 
 \frac{ e^{ i ( {\bf{q}} \cdot {\bf{r}} - \tilde{\omega}_{n} \tau )}}{ i \tilde{\omega}_{n}
 -  {\bf{v}}^{\alpha} \cdot {\bf{q}} } 
 \; \; \; ,
 \label{eq:G0res}
 \\
 \Phi^{\alpha} ( {\bf{r}} , \tau ) & = &
  \sum_{q} 
 \frac{ e^{ i ( {\bf{q}} \cdot {\bf{r}} - \omega_{m} \tau )}}{ i \omega_{m}
 -  {\bf{v}}^{\alpha} \cdot {\bf{q}} } V^{\alpha}_{q}
 \; \; \; .
 \label{eq:Phires}
 \end{eqnarray}
Note that the bosonic Matsubara frequencies in Eq.\ref{eq:Phires} insure
that $\Phi^{\alpha} ( {\bf{r}} , \tau + \beta )  = 
\Phi^{\alpha} ( {\bf{r}} , \tau )  $, so that
the ansatz \ref{eq:Ansatz} manifestly satisfies the KMS boundary 
condition \ref{eq:KMS}.
Gaussian averaging of Eq.\ref{eq:Ansatz} with the effective action
$S_{eff} \{ \phi^{\alpha} \}$
given in Eqs.\ref{eq:S2decoupdef} and 
\ref{eq:Seff2phigaussres} yields
 \begin{equation}
\left<  {\cal{G}}^{\alpha} ( {\bf{r}} , {\bf{r}}^{\prime} , \tau , \tau^{\prime} )
\right>_{S_{eff}} =
 {{G}}^{\alpha}_{0} ( {\bf{r}} - {\bf{r}}^{\prime} , \tau - \tau^{\prime} )
 e^{Q^{\alpha}
 ( {\bf{r}} - {\bf{r}}^{\prime} , \tau - \tau^{\prime} )}
 \label{eq:bosres}
 \; \; \; ,
 \end{equation}
 \begin{equation}
 Q^{\alpha} ( {\bf{r}} , \tau ) =
 \frac{1}{\beta V} \sum_q
 \frac{ f^{RPA , \alpha}_q }{ ( i \omega_m - {\bf{v}}^{\alpha}
 \cdot {\bf{q}} )^2 }
 \left[ 1 - \cos ( {\bf{q}} \cdot {\bf{r}} - \omega_m \tau ) \right]
 \label{eq:DebyeWaller}
 \; \; \; .
 \end{equation}
Here $f^{RPA , \alpha}_q$  
is the screened interaction within random-phase approximation (RPA),
\begin{equation}
 f^{RPA , \alpha}_q  =
 \left[ \underline{f}_q \left[
 1 + \underline{\Pi}_0 ( q ) \underline{f}_q \right]^{-1}
 \right]^{\alpha \alpha}
 \; \; \;  ,
 \end{equation}
where underlined quantities are matrices in the patch indices.
For the special case of the one-dimensional
Tomonaga-Luttinger model\cite{Tomonaga50} Eq.\ref{eq:bosres} {\it{exactly}}
reproduces the well-known bosonization result\cite{Fogedby76,Kopietzhab}.

\section{The eikonal expansion}
\label{sec:eik}

An obvious disadvantage of the procedure outlined in the previous section is
that the Schwinger solution in Eqs.\ref{eq:Ansatz}-\ref{eq:Phires}
is only valid for linearized energy dispersion.
In general, the effective energy dispersion in box 
$K^{\alpha}$ is of the form
$ \epsilon_{ {\bf{k}}^{\alpha} + {\bf{q}}} = \epsilon_{ {\bf{k}}^\alpha} + {\bf{v}}^{\alpha} \cdot {\bf{q}}
 + \frac{ {\bf{q}}^2}{2 m^{\alpha}}$.
The eikonal method can in principle handle the quadratic 
(curvature) term in a systematic way.
However, this method cannot be directly compared with our functional
bosonization approach, because it is by construction a {\it{real time}} method.
Let us briefly summarize the main steps. 
For simplicity we focus here on the
{\it{retarded}} Green's function
for fixed {\it{real time}} background field
$\tilde{V}^{\alpha}  ( {\bf{r}} , t )$,
which is defined via the differential equation
 \begin{equation}
  \left[i {\partial}_{ t} -    
  {\xi}^{\alpha} ( {\bf{P}}_{\bf{r}} ) - \tilde{V}^{\alpha}  ( {\bf{r}} , t )
  \right]
 {\cal{G}}^{\alpha}_{ret} ( {\bf{r}} , {\bf{r}}^{\prime} , t , t^{\prime} )
 = \delta ( {\bf{r}} - {\bf{r}}^{\prime} ) \delta ( t - t^{\prime} )
 \; \; \; ,
 \label{eq:Galpharetdifrt}
 \end{equation}
with the boundary condition that 
$ {\cal{G}}^{\alpha}_{ret} ( {\bf{r}} , {\bf{r}}^{\prime} , t , t^{\prime} ) = 0$ for
$t - t^{\prime} < 0$.
In Sec.\ref{sec:Mink} we shall discuss how to obtain for the
leading term in the eikonal expansion the
corresponding {\it{time-ordered}} Green's function,
which is usually calculated
in diagrammatic perturbation theory.
The eikonal expansion is a
systematic method for solving partial differential equations of the type
\ref{eq:Galpharetdifrt}.  This approach
has been developed many years ago\cite{Svidzinskii57,Barbashov65,Fradkin66,Popov83}.
In this section we shall briefly outline how
Eq.\ref{eq:Galpharetdifrt} can be solved with this method, following mainly 
Ref.\cite{Fradkin66}.

It is instructive to study first the solution of Eq.\ref{eq:Galpharetdifrt}
without external potential, i.e. for $\tilde{V}^{\alpha} ( {\bf{r}}, t ) = 0$.
Then the retarded propagator is given by\cite{footnoteG}
 \begin{equation}
 {{G}}^{\alpha}_{ret,0} ( {\bf{r}} - {\bf{r}}^{\prime} , t - t^{\prime} )
 = \int \frac{d {\bf{q}}}{ ( 2 \pi )^d} \int \frac{d \omega}{ 2 \pi}
 e^{ i [ {\bf{q}} \cdot ( {\bf{r}} - {\bf{r}}^{\prime} ) - \omega ( t - t^{\prime} ) ] }
 {{G}}^{\alpha}_{ret,0} ( {\bf{q}} , \omega )
 \label{eq:Gret0FT}
 \; \; \; ,
 \end{equation}
with
 \begin{equation}
 {{G}}^{\alpha}_{ret,0} ( {\bf{q}} , \omega )
  =  \frac{1}{ \omega - \xi^{\alpha} ( {\bf{q}} ) + i 0^{+} }
  =  - i \int_0^{\infty} d t^{\prime \prime} e^{i t^{\prime \prime} [ 
 \omega - \xi^{\alpha} ( {\bf{q}} ) + i 0^{+} ] }
 \label{eq:Gret0qo}
 \; \; \; .
 \end{equation}
As will be evident shortly,
the representation of the propagator in terms of an integral over an auxiliary 
time $t^{\prime \prime}$ plays an important role in the eikonal approach.
Note, however, that we can exactly eliminate the auxiliary integral
by substituting Eq.\ref{eq:Gret0qo} into Eq.\ref{eq:Gret0FT} and doing first
the frequency integration, using
 $\int \frac{d \omega}{ 2 \pi} e^{ -i \omega ( t-t^{\prime} - t^{\prime \prime} ) }
 = \delta ( t - t^{\prime} - t^{\prime \prime} )$.
Then the $t^{\prime \prime}$-integral is trivial, and we obtain
 \begin{equation}
 {{G}}^{\alpha}_{ret,0} ( {\bf{r}} - {\bf{r}}^{\prime} , t - t^{\prime} )
 = - i \Theta ( t - t^{\prime} ) \int \frac{d {\bf{q}}}{ ( 2 \pi )^d} 
 e^{ i [ {\bf{q}} \cdot ( {\bf{r}} - {\bf{r}}^{\prime} ) - \xi^{\alpha} ( 
 {\bf{q}} ) ( t - t^{\prime} ) ] }
 \label{eq:Gret0FT1}
 \; \; \; .
 \end{equation}
If we linearize the energy dispersion (i.e. set $m^{\alpha} = \infty$,
as in conventional bosonization),
then the ${\bf{q}}$-integration can be performed exactly, with the result
 \begin{equation}
 {{G}}^{\alpha}_{ret,0} ( {\bf{r}} - {\bf{r}}^{\prime} , t - t^{\prime} )
 = - i \Theta ( t - t^{\prime} ) \delta^{(d)} (
  {\bf{r}} - {\bf{r}}^{\prime} - {\bf{v}}^{\alpha} ( t - t^{\prime}) )
  \; \; \; , \; \; \; m^{\alpha} =  \infty
  \; \; \; .
  \label{eq:G0retdelta}
  \end{equation}

Let us now consider the case $\tilde{V}^{\alpha} 
( {\bf{r}}, t ) \neq 0$ . 
The starting point is the transformation
of Eq.\ref{eq:Galpharetdifrt} to the mixed representation 
by Fourier transforming it with  respect to the difference variables
$ {\bf{r}} - {\bf{r}}^{\prime} $ and $t - t^{\prime}$\cite{Fradkin66}. Defining
 \begin{equation}
 {\cal{G}}^{\alpha}_{ret} ( {\bf{r}} , {\bf{r}}^{\prime} , t , t^{\prime} )
 = \int \frac{d {\bf{q}}}{ ( 2 \pi )^d} \int \frac{d \omega}{ 2 \pi}
 e^{ i [ {\bf{q}} \cdot ( {\bf{r}} - {\bf{r}}^{\prime} ) - \omega ( t - t^{\prime} ) ] }
 {\cal{G}}^{\alpha}_{ret} ( {\bf{r}}, t ; {\bf{q}} , \omega )
 \label{eq:Gretmixed}
 \; \; \; ,
 \end{equation}
it is easy to see that Eq.\ref{eq:Galpharetdifrt} reduces to
 \begin{equation}
  \left[i {\partial}_{ t} -    
  {\xi}^{\alpha} ( {\bf{P}}_{\bf{r}} ) - \tilde{V}^{\alpha}  ( {\bf{r}} , t )
  + \omega - \xi^{\alpha} ( {\bf{q}} )
  \right]
 {\cal{G}}^{\alpha}_{ret} ( {\bf{r}} , t ; {\bf{q}} , \omega )
 = 1
 \; \; \; .
 \label{eq:Galphartqo}
 \end{equation}
The formal solution of this equation can be written as\cite{Fradkin66}
 \begin{equation}
 {\cal{G}}^{\alpha}_{ret} ( {\bf{r}} , t ; {\bf{q}} , \omega )
 =
 - i \int_0^{\infty} d t^{\prime \prime} e^{i t^{\prime \prime} [ 
 \omega - \xi^{\alpha} ( {\bf{q}} ) + i 0^{+} ] }
 Y^{\alpha} ( t^{\prime \prime} ; {\bf{r}} , t )
 \label{eq:Gformal}
 \; \; \; ,
 \end{equation}
 where the auxiliary function 
 $Y^{\alpha} ( t^{\prime \prime} ; {\bf{r}} , t )$ satisfies the partial differential
 equation
  \begin{equation}
  \left[ i \partial_{t^{\prime \prime} } + i {\partial}_{ t} -    
  {\xi}^{\alpha} ( {\bf{P}}_{\bf{r}} ) - \tilde{V}^{\alpha}  ( {\bf{r}} , t )
  \right]
  Y^{\alpha} ( t^{\prime \prime} ; {\bf{r}} , t ) = 0
  \label{eq:Ydef}
  \; \; \; ,
  \end{equation}
with boundary condition
$  Y^{\alpha} ( 0 ; {\bf{r}} , t ) = 1$.
The correctness of Eq.\ref{eq:Gformal} is easily verified by
applying the differential operator in Eq.\ref{eq:Galphartqo} to the
right-hand side of Eq.\ref{eq:Gformal} and integrating by parts.
Note also that without external potential the unique solution
of Eq.\ref{eq:Ydef} with the correct boundary condition is
$Y^{\alpha} ( t^{\prime \prime} ; {\bf{r}} , t ) = 1$, so that
Eq.\ref{eq:Gformal} reduces to the right-hand side of Eq.\ref{eq:Gret0qo}.
It is convenient to parametrize the solution of Eq.\ref{eq:Ydef} in the
form
$  Y^{\alpha} ( t^{\prime \prime} ; {\bf{r}} , t ) = 
  e^{F^{\alpha} ( t^{\prime \prime} ; {\bf{r}} , t ) }$.
The equivalent differential equation for $F^{\alpha}$ is then
  \begin{equation}
  \left[ i \partial_{t^{\prime \prime} } + i {\partial}_{ t} -    
  {\xi}^{\alpha} ( {\bf{P}}_{\bf{r}} ) 
  \right]
  F^{\alpha} ( t^{\prime \prime} ; {\bf{r}} , t ) = 
   \tilde{V}^{\alpha}  ( {\bf{r}} , t )
   + \frac{[ {\bf{P}}_{\bf{r}} 
  F^{\alpha} ( t^{\prime \prime} ; {\bf{r}} , t ) ]^2 }{2 m^{\alpha}}
  \label{eq:Fdif}
  \; \; \; ,
  \end{equation}
with boundary condition
$  F^{\alpha} ( 0 ; {\bf{r}} , t ) = 0$.
This non-linear partial differential equation cannot be solved exactly.
However, we can obtain the solution as expansion
in powers of the external potential $\tilde{V}^{\alpha}$\cite{Fradkin66}.
Substituting the ansatz
$  F^{\alpha} ( t^{\prime \prime} ; {\bf{r}} , t ) = 
  \sum_{n=1}^{\infty} 
  F^{\alpha}_n ( t^{\prime \prime} ; {\bf{r}} , t )  $
into Eq.\ref{eq:Fdif} (where $F^{\alpha}_n$ is by 
assumption of order $( \tilde{V}^{\alpha} )^n$), we have
  \begin{equation}
  \left[ i \partial_{t^{\prime \prime} } + i {\partial}_{ t} -    
  {\xi}^{\alpha} ( {\bf{P}}_{\bf{r}} ) 
  \right]
  F^{\alpha}_n ( t^{\prime \prime} ; {\bf{r}} , t ) = 
   \tilde{V}^{\alpha}_{n}  ( t^{\prime \prime} ; {\bf{r}} , t )
  \label{eq:Fndif}
  \; \; \; ,
  \; \; \; n = 1,2, \ldots
  \; \; \; ,
  \end{equation}
with $\tilde{V}^{\alpha}_{1}  ( t^{\prime \prime} ; {\bf{r}} , t )=
\tilde{V}^{\alpha}  ( {\bf{r}} , t )$ (independent of $t^{\prime \prime}$), and
   \begin{equation}
   \tilde{V}^{\alpha}_{n}  ( t^{\prime \prime} ; {\bf{r}} , t )
    =  \frac{1}{2 m^{\alpha} } \sum_{n^{\prime}=1}^{n-1}
   [ {\bf{P}}_{\bf{r}} 
  F^{\alpha}_{n^{\prime}} ( t^{\prime \prime} ; {\bf{r}} , t )  ] \cdot
   [ {\bf{P}}_{\bf{r}} 
  F^{\alpha}_{n-n^{\prime}} ( t^{\prime \prime} ; {\bf{r}} , t )  ] 
  \; \; \; ,
  \; \; \; n = 2,3, \ldots
  \; \; \; .
  \end{equation}
Eq.\ref{eq:Fndif} is easily solved by means of the Green's function of the
differential operator on the left-hand side. The solution with the
correct boundary condition can be written as
  \begin{equation}
  F^{\alpha}_n ( t^{\prime \prime} ; {\bf{r}} , t ) = 
  - i \int_{0}^{ t^{\prime \prime} } d s
  e^{i [ i \partial_t - \xi^{\alpha} ( {\bf{P}}_{\bf{r}} ) ] ( t^{\prime \prime} - s ) }
  \tilde{V}^{\alpha}_{n}  ( s ; {\bf{r}} , t )
  \label{eq:Fnsol}
  \; \; \; .
  \end{equation}
This completes the formal solution of the problem.

\section{Formal similarities between bosonization 
and the eikonal expansion for linearized energy dispersion}
\label{sec:equivalence}

Suppose we truncate the eikonal expansion at the first order,
$  F^{\alpha} ( t^{\prime \prime} ; {\bf{r}} , t ) \approx
  F^{\alpha}_1 ( t^{\prime \prime} ; {\bf{r}} , t )  $.
This approximation
becomes exact in the limit $m^{\alpha} = \infty$
(i.e. for linearized energy dispersion), because then
all higher order potentials 
$\tilde{V}^{\alpha}_{n}$ with $n \geq 2$ in Eq.\ref{eq:Fndif} vanish identically.
For finite $m^{\alpha}$, we expect that the relevant small
parameter which controls this approximation is
proportional to the
{\it{product}} of $1/m^{\alpha}$ times the typical strength of the effective screened
interaction\cite{Kopietz95,Kopietzhab}. 
The function $F^{\alpha}_1 ( t^{\prime \prime} ; {\bf{r}} , t )  $ is easily calculated by
decomposing $\tilde{V}^{\alpha}  ( {\bf{r}} , t )$ into its Fourier components,
 \begin{equation}
 \tilde{V}^{\alpha}  ( {\bf{r}} , t ) =  
  \int \frac{d {\bf{q}}}{ ( 2 \pi )^d} \int \frac{d \omega}{ 2 \pi}
 e^{ i [ {\bf{q}} \cdot  {\bf{r}}  - \omega  t  ] }
 \tilde{V}^{\alpha} ( {\bf{q}}, \omega )
 \; \; \; ,
 \end{equation}
and using
  \begin{equation}
  e^{i [ i \partial_t - \xi^{\alpha} ( {\bf{P}}_{\bf{r}} ) ] ( t^{\prime \prime} - s ) }
 e^{ i [ {\bf{q}} \cdot  {\bf{r}}  - \omega  t  ] }
 = 
  e^{i [ \omega - \xi^{\alpha} ( {\bf{q}} ) ] ( t^{\prime \prime} - s ) }
 e^{ i [ {\bf{q}} \cdot  {\bf{r}}  - \omega  t  ] }
 \; \; \; .
 \end{equation}
After performing the $s$-integration in Eq.\ref{eq:Fnsol} we obtain
  \begin{equation}
  F^{\alpha}_1 ( t^{\prime \prime} ; {\bf{r}} , t )  
  =
  \int \frac{d {\bf{q}}}{ ( 2 \pi )^d} \int \frac{d \omega}{ 2 \pi}
 e^{ i [ {\bf{q}} \cdot  {\bf{r}}  - \omega  t  ] }
 \frac{ \tilde{V}^{\alpha} ( {\bf{q}}, \omega )}
 { \omega - \xi^{\alpha} ( {\bf{q}} ) }
 \left[ 1 - e^{ i [ \omega - \xi^{\alpha} ( {\bf{q}} ) ] t^{\prime \prime}} \right]
 \; \; \; .
 \label{eq:F1sol}
 \end{equation}
The leading eikonal result for the retarded Green's function in a given
external potential is then 
 \begin{eqnarray}
 {\cal{G}}^{\alpha}_{ret,1} ( {\bf{r}} , {\bf{r}}^{\prime} , t , t^{\prime} )
 & = & \int \frac{d {\bf{q}}}{ ( 2 \pi )^d} \int \frac{d \omega}{ 2 \pi}
 e^{ i [ {\bf{q}} \cdot ( {\bf{r}} - {\bf{r}}^{\prime} ) - \omega ( t - t^{\prime} ) ] }
 {\cal{G}}^{\alpha}_{ret,1} ( {\bf{r}}, t ; {\bf{q}} , \omega )
 \label{eq:Gret1FT}
 \; \; \; ,
 \\
 {\cal{G}}^{\alpha}_{ret,1} ( {\bf{r}}, t ; {\bf{q}} , \omega )
 & = & - i \int_0^{\infty} d t^{\prime \prime} e^{i t^{\prime \prime} [ 
 \omega - \xi^{\alpha} ( {\bf{q}} ) + i 0^{+} ] }
 e^{
  F^{\alpha}_1 ( t^{\prime \prime} ; {\bf{r}} , t )  }
 \label{eq:Gret1mixed}
 \; \; \; .
 \end{eqnarray}
We now show that {\it{for linearized energy dispersion}}
(i.e. for $m^{\alpha}= \infty$) Eqs.\ref{eq:Gret1FT}
and \ref{eq:Gret1mixed} have {\it{exactly}} the same structure as the
Schwinger solution given in Eqs.\ref{eq:Ansatz}-\ref{eq:Phires}.
At the first sight this is not at all clear, because the eikonal result
in Eqs.\ref{eq:Gret1FT} and \ref{eq:Gret1mixed}
involves an  additional integration over the auxiliary time
$t^{\prime \prime}$. In the work of Khveshchenko and Stamp\cite{Khveshchenko93}
the $t^{\prime \prime}$-integration has been performed by means of the saddle-point
method. However, to see the connection with bosonization, we
eliminate the auxiliary variable
$t^{\prime \prime}$ from Eq.\ref{eq:Gret1mixed}. 
Therefore we simply repeat the manipulations 
leading to  Eq.\ref{eq:Gret0FT1}  and first perform the $\omega$-integration in
Eq.\ref{eq:Gret1FT}. As before, this gives rise to a factor of
$\delta ( t - t^{\prime} -  t^{\prime \prime} )$.
The $t^{\prime \prime}$-integration is then trivial, and we obtain
 \begin{equation}
 {\cal{G}}^{\alpha}_{ret,1} ( {\bf{r}} , {\bf{r}}^{\prime} , t , t^{\prime} )
 =
 {{G}}^{\alpha}_{ret,0} ( {\bf{r}} - {\bf{r}}^{\prime} , t - t^{\prime} )
 e^{
  F^{\alpha}_1 ( t- t^{\prime} ; {\bf{r}} , t )  }
 \; \; \; ,
 \label{eq:Gret1int}
 \end{equation}
with
 ${{G}}^{\alpha}_{ret,0} ( {\bf{r}} - {\bf{r}}^{\prime} , t - t^{\prime} )$ given
in Eq.\ref{eq:Gret0FT1}.
Still, the exponential factor in Eq.\ref{eq:Gret1int} does
not have the same structure as the corresponding factor
in Eq.\ref{eq:Ansatz}.
However, for $m^{\alpha} = \infty$ the function
 ${{G}}^{\alpha}_{ret,0} ( {\bf{r}} - {\bf{r}}^{\prime} , t - t^{\prime} )$ 
is proportional to
 $ \delta^{(d)} (
  {\bf{r}} - {\bf{r}}^{\prime} - {\bf{v}}^{\alpha} ( t - t^{\prime}) )$
(see Eq.\ref{eq:G0retdelta}), so that we may replace
${\bf{v}}^{\alpha} ( t - t^{\prime})  \rightarrow
{\bf{r}} - {\bf{r}}^{\prime} $
in the expression for
$F^{\alpha}_1 ( t- t^{\prime} ; {\bf{r}} , t )$ 
in Eq.\ref{eq:Gret1int},
 \begin{eqnarray}
 e^{ i [ {\bf{q}} \cdot  {\bf{r}}  - \omega  t  ] }
 \left[
  1 - e^{ i [ \omega - \xi^{\alpha} ( {\bf{q}} ) ] (t - t^{\prime} ) }  \right]
  & = &
 e^{ i [ {\bf{q}} \cdot  {\bf{r}}  - \omega  t  ] }
 \left[
  1 -
  e^{ i  \omega (t - t^{\prime}) - i {\bf{q}} \cdot {\bf{v}}^{\alpha} (t - t^{\prime} ) } 
  \right]
  \nonumber
  \\
   \rightarrow 
 e^{ i [ {\bf{q}} \cdot  {\bf{r}}  - \omega  t  ] }
 \left[
  1 -
  e^{ i  [ \omega (t - t^{\prime}) -  {\bf{q}} \cdot ( {\bf{r}} - {\bf{r}}^{\prime}  ) ]} 
  \right]
  & = &
 e^{ i [ {\bf{q}} \cdot  {\bf{r}}  - \omega  t  ] }
 -
 e^{ i [ {\bf{q}} \cdot  {\bf{r}}^{\prime}  - \omega  t^{\prime}  ] }
  \; \; \; .
  \end{eqnarray}
Then it is easy to see that Eq.\ref{eq:Gret1int}
can also be written as
 \begin{equation}
 {\cal{G}}^{\alpha}_{ret,1} ( {\bf{r}} , {\bf{r}}^{\prime} , t , t^{\prime} )
 =
 {{G}}^{\alpha}_{ret,0} ( {\bf{r}} - {\bf{r}}^{\prime} , t - t^{\prime} )
 e^{
  \Phi^{\alpha}_1 ( {\bf{r}} , t )   -
  \Phi^{\alpha}_1 ( {\bf{r}}^{\prime} , t^{\prime} )  } 
 \; \; \; ,
 \label{eq:Gret2int}
 \end{equation}
with
  \begin{equation}
  \Phi^{\alpha}_1 ( {\bf{r}} , t )   =
  \int \frac{d {\bf{q}}}{ ( 2 \pi )^d} \int \frac{d \omega}{ 2 \pi}
 e^{ i [ {\bf{q}} \cdot  {\bf{r}}  - \omega  t  ] }
 \frac{ \tilde{V}^{\alpha} ( {\bf{q}}, \omega )}
 { \omega - {\bf{v}}^{\alpha} \cdot {\bf{q}}  }
 \; \; \; .
 \label{eq:Phi1sol}
 \end{equation}
The formal similarity to Eqs.\ref{eq:Ansatz}-\ref{eq:Phires} is now evident.
Note, however, that the imaginary time finite-temperature Green's function
discussed in Sec.\ref{sec:bos} and the
real time retarded Green's function discussed here are not related
in a trivial way: only after Fourier transformation
into the frequency domain there exists a simple relation via the spectral
function.

\section{Averaging in Minkowski time: How good is the eikonal expansion for
linearized energy dispersion?}
\label{sec:Mink}

In spite of the formal similarity between the real time eikonal result
in Eq.\ref{eq:Gret2int}
and the imaginary time bosonization solution in Eq.\ref{eq:Ansatz}, 
after the averaging procedure the final expressions for the Green's function
of the many-body system are {\it{not}} equivalent.
To average  Eq.\ref{eq:Gret1mixed}, one should keep in mind
that by construction functional
averaging always generates {\it{time-ordered}} correlation
functions\cite{ZinnJustin89}.
Hence, in order to properly define the averaging procedure, we
should replace the retarded Green's function in
Eq.\ref{eq:Gret1mixed} 
by the corresponding time-ordered Green's function {\it{before}}
averaging with respect to the effective action
$\tilde{S}_{eff}$ of the background field.
This is easily done by going back
to Eq.\ref{eq:Gret0qo} and choosing the limit for the 
$t^{\prime \prime}$-integration such that the integral for the
time-ordered Green's function is convergent.
With the notation $s_{\omega} = \mbox{sign} \omega$ 
we have
 \begin{equation}
 {{G}}^{\alpha}_{0} ( {\bf{q}} , \omega )
  =  \frac{1}{ \omega - \xi^{\alpha} ( {\bf{q}} ) + i 0^{+} s_{\omega} }
  =  - i \int_0^{\infty s_\omega } d 
  t^{\prime \prime} e^{i t^{\prime \prime} [ 
 \omega - \xi^{\alpha} ( {\bf{q}} ) + i 0^{+} s_\omega ] }
 \label{eq:G0qo}
 \; \; \; .
 \end{equation}
Using the fact that functional averaging
restores translational invariance in space and time, 
it is easy to see that after averaging the
leading eikonal result for the {\it{time-ordered}} Green's function reads
 \begin{equation}
 {{G}}^{\alpha}_{1} ( {\bf{q}} , \omega )
  =  - i \int_0^{\infty s_\omega } d t^{\prime \prime} e^{i t^{\prime \prime} [ 
 \omega - \xi^{\alpha} ( {\bf{q}} ) + i 0^{+} s_\omega  ] }
 \left<
 e^{
  F^{\alpha}_1 ( t^{\prime \prime} ; 0,0 )  }
  \right>_{\tilde{S}_{eff}}
 \label{eq:Gret1mixedav}
 \; \; \; .
 \end{equation}
At the level of the Gaussian approximation the functional average in Eq.\ref{eq:Gret1mixedav}
generates the usual Debye-Waller factor, with propagator given by the
RPA-interaction\cite{Khveshchenko93}. Thus,
 \begin{equation}
 {{G}}^{\alpha}_{1} ( {\bf{q}} , \omega )
  =  - i \int_0^{\infty s_\omega } d t^{\prime \prime} e^{i t^{\prime \prime} [ 
 \omega - \xi^{\alpha} ( {\bf{q}} ) + i 0^{+} s_\omega ] }
 e^{
  Q^{\alpha}_1 ( t^{\prime \prime} )  }
 \label{eq:Qret}
 \; \; \; ,
 \end{equation}
with
 \begin{equation}
  Q^{\alpha}_1 ( t^{\prime \prime} )   =
  \int \frac{d {\bf{q}}^{\prime} }{ ( 2 \pi )^d} \int \frac{d \omega^{\prime}}{ 2 \pi i}
   f^{RPA} ( {\bf{q}}^{\prime} , \omega^{\prime} ) 
 \frac{  1 - 
 \cos \left[ ( \omega^{\prime} - \xi^{\alpha} ({\bf{q}}^{\prime}) ) t^{\prime \prime} \right]  }
  { [ \omega^{\prime} - \xi^{\alpha} ( {\bf{q}}^{\prime} ) ]^2 }
 \label{eq:Qretres}
 \; \; \; .
 \end{equation}
For simplicity we have assumed a patch-- and frequency-independent bare interaction
$f_{\bf{q}}$, so that the 
screened interaction can be identified with the usual
RPA interaction,
   $f^{RPA} ( {\bf{q}} , \omega ) 
   = f_{\bf{q}}[{1 + \Pi_0 ( {\bf{q}} , \omega ) f_{\bf{q}} }]^{-1}$,
where $\Pi_0 ( {\bf{q}} , \omega )$ is the Lindhard function\cite{Fetter71}.
In Eq.\ref{eq:Qretres} we have called the integration 
variables ${\bf{q}}^{\prime}$ and $\omega^{\prime}$
to avoid confusion with the external labels ${\bf{q}}$ and $\omega$ in
Eq.\ref{eq:Qret}.

According to Eq.\ref{eq:Qret} the spectral function
$ - \frac{1}{\pi} Im {{G}}^{\alpha}_{1} ( {\bf{q}} , \omega + i 0^{+})$ of our interacting
many-body system can be obtained via a {\it{one-dimensional}}
Fourier transformation. Furthermore, 
except for the trivial factor of $s_{\omega}$,
the spectral function depends 
exclusively on the combination
$\omega - \xi^{\alpha} ( {\bf{q}} )$.
In particular, for linearized energy dispersion $\xi^{\alpha}
( {\bf{q}} ) = {\bf{v}}^{\alpha} \cdot {\bf{q}}$ (where the eikonal
expansion truncates at the first order,
so that Eq.\ref{eq:Gret2int} is the exact solution of
Eq.\ref{eq:Galpharetdifrt}) the spectral
function is according to Eq.\ref{eq:Qret} a combination of the variable 
$\omega - {\bf{v}}^{\alpha} \cdot {\bf{q}}$ only.
The well-known bosonization result for the Green's function
of the Tomonaga-Luttinger model\cite{Tomonaga50} 
(i.e. one-dimensional electrons
with long-range interactions and exactly linear energy dispersion)
shows that in one dimension 
this result cannot be correct\cite{Dzyaloshinskii74}!
The spectral function of the Tomonaga-Luttinger model
exhibits a more complicated dependence on  ${\bf{q}}$ and $\omega$\cite{Meden92}.
One possibility to cure this 
problem might be to
choose some effective $\omega$-- or  ${\bf{q}}$-dependent
cutoffs for the integration limits in Eq.\ref{eq:Qretres}\cite{Khveshchenkopriv}.
However, such a procedure seems to require a certain amount of 
intuition and knowledge about the 
final result,
and is certainly not satisfactory from the formal point of view.

\begin{figure}
\vspace{1cm}
\hspace{2cm}
\epsfysize7cm 
\epsfbox{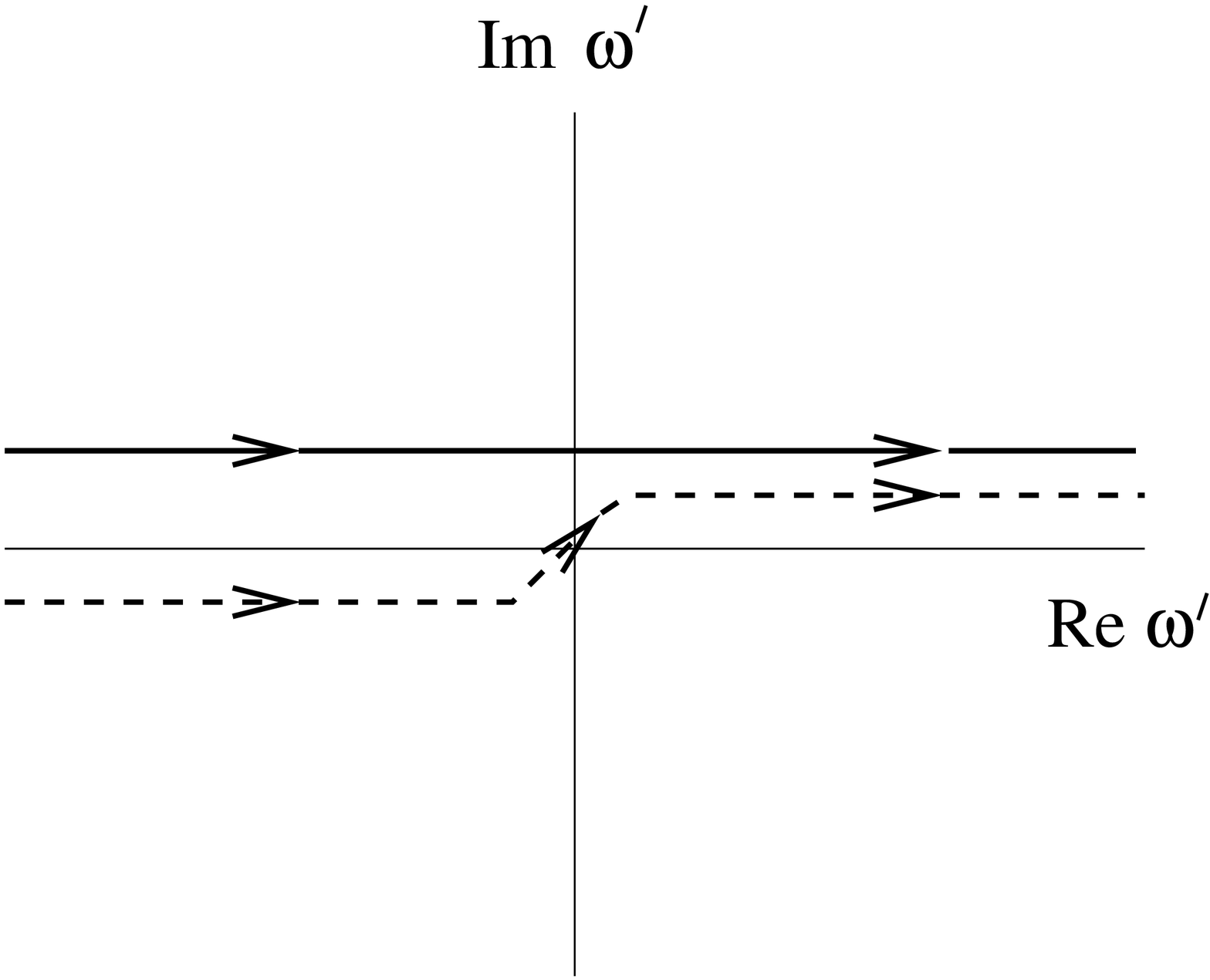}
\vspace{1cm}
\caption{Possible integration contours in the complex $\omega^{\prime}$-plane
for the evaluation of Eq.\ref{eq:Qretres}.
The dashed  line is the correct time-ordered contour. 
The solid line corresponds to choosing the
retarded interaction in Eq.\ref{eq:Qretres}. Both contours are 
infinitesimally close to the real axis.}
\label{fig:contour}
\end{figure}

The obvious question is now whether the
spectral function calculated from the
real time eikonal expansion given in
Eqs.\ref{eq:Qret} and \ref{eq:Qretres}
agrees at least approximately with the 
spectral function calculated from the imaginary time bosonization
result given in Eqs.\ref{eq:bosres} and \ref{eq:DebyeWaller}.
To clarify this point, it is useful to rewrite Eq.\ref{eq:Qretres}
with the help of the dynamic structure factor\cite{Pines89}, which
is related to the RPA interaction via\cite{Kopietzhab}
 \begin{equation}
 f^{RPA} ( {\bf{q}}^{\prime} , \omega^{\prime} ) = f_{\bf{q}^{\prime}}
 - f_{\bf{q}^{\prime}}^2 \int_{0}^{\infty} d \omega
 S^{RPA} ( {\bf{q}}^{\prime} , \omega ) 
 \left[
 \frac{ 1} {  \omega - \omega^{\prime} }
 + \frac{ 1} {  \omega + \omega^{\prime} } \right]
 \label{eq:fspec}
\; \; \; .
\end{equation}
Substituting this expression into Eq.\ref{eq:Qretres}, it is obvious
that we encounter poles on the real $\omega^{\prime}$-axis, which must
be regularized by an appropriate deformation of the integration contour in
the complex $\omega^{\prime}$-plane. 
The correct contour is easily determined from the requirement that
the expansion of the factor $e^{ Q^{\alpha}_1 ( t^{\prime \prime} ) }$ in
Eq.\ref{eq:Qret} should reproduce the lowest order  
perturbation theory for the {\it{time-ordered}} Green's function\cite{Quinn58}.
It is easy to see that this implies that we
use the dashed contour shown in Fig.\ref{fig:contour}, 
which describes time-ordering
and corresponds to the regularization
$\omega^{\prime} \rightarrow \omega^{\prime} + i 0^{+} \mbox{sign} 
\omega^{\prime}$ in Eq.\ref{eq:fspec}.
The $\omega^{\prime}$-integral can then be carried out via
contour integration. We obtain
$Q_1^{\alpha} ( t^{\prime \prime} ) = R_1^{\alpha} - S_1^{\alpha} ( t^{\prime \prime} )$,
with
 \begin{equation}
 R^{\alpha}_1 = -  \int  
 \frac{d {\bf{q}}}{ (2 \pi)^d} f_{\bf{q}}^2
 \int_{0}^{\infty} d \omega \frac{S^{RPA} ( {\bf{q}} , \omega )}{ ( \omega + | {\bf{v}}^{\alpha}
 \cdot {\bf{q}} | )^2}
 \label{eq:R2}
 \; \; \; ,
 \end{equation}
 \begin{eqnarray}
 S^{\alpha}_1 ( t^{\prime \prime} ) & = & 
 \frac{i t^{\prime \prime}}{2}
 \int
 \frac{d {\bf{q}}}{ (2 \pi)^d} f^{RPA} ( {\bf{q}} , {\bf{v}}^{\alpha} \cdot {\bf{q}} )
 -  \int  
 \frac{d {\bf{q}}}{ (2 \pi)^d} f_{\bf{q}}^2
 \int_{0}^{\infty} d \omega \frac{S^{RPA} ( {\bf{q}} , \omega )}{ ( \omega + | {\bf{v}}^{\alpha}
 \cdot {\bf{q}} | )^2}
 \nonumber
 \\
 &  & \hspace{-15mm} \times \left\{
 \frac{
 e^{- i \omega t^{\prime \prime} } \left[ ( \omega^2 + ( {\bf{v}}^{\alpha} \cdot {\bf{q}} )^2 )
 \cos ( {\bf{v}}^{\alpha} \cdot {\bf{q}} t^{\prime \prime} )
 + 2 i \omega {\bf{v}}^{\alpha} \cdot {\bf{q}} 
 \sin ( {\bf{v}}^{\alpha} \cdot {\bf{q}} t^{\prime \prime} ) \right]
 - 2 \omega | {\bf{v}}^{\alpha} \cdot {\bf{q}} | }
 {
 ( \omega - | {\bf{v}}^{\alpha}
 \cdot {\bf{q}} | )^2}
 \right\}
 \label{eq:S2}
 \; \; \; .
 \end{eqnarray}
It is easy to show that the term $R^{\alpha}_1$ agrees exactly with the
zero-temperature limit of the space-- and time-independent
contribution to $Q^{\alpha} ( {\bf{r}} , \tau )$ in
Eq.\ref{eq:DebyeWaller},
\begin{equation}
R^{\alpha}_1 = \lim_{ \beta \rightarrow \infty}
\lim_{V \rightarrow \infty}
 \frac{1}{\beta V} \sum_q
 \frac{ f^{RPA , \alpha}_q }{ ( i \omega_m - {\bf{v}}^{\alpha}
 \cdot {\bf{q}} )^2 }
 \label{eq:Ragree}
 \; \; \; .
 \end{equation}
Because in a Fermi liquid the quantity $e^{R^{\alpha}_1}$ can be identified
with the quasi-particle residue\cite{Kopietz94,Kopietzhab}, it is clear that for
Fermi liquids the real time eikonal approach
and functional bosonization yield exactly the same non-perturbative 
result for the quasi-particle residue.
The proper deformation of the contour into the complex $\omega^{\prime}$-plane
was crucial to obtain this result. If we had 
directly averaged the eikonal result for the
retarded Green's function in Eq.\ref{eq:Gret1mixed} 
(choosing the retarded contour shown in Fig.\ref{fig:contour}), we would have obtained
incorrectly $R^{\alpha}_1 = 0$.

The time-dependent contribution $S^{\alpha}_1 ( t^{\prime \prime} )$
cannot be directly compared with a corresponding term in Eq.\ref{eq:DebyeWaller}.
To  examine possible
discrepancies with bosonization, let us explicitly calculate 
$Q^{\alpha}_1 ( t^{\prime \prime} )$
for the spinless
Tomonaga-Luttinger model\cite{Tomonaga50} with interaction parameters
$g_4 ( {\bf{q}} ) = g_2 ( {\bf{q}} ) =f_{\bf{q}}$. 
In this case the patch index $\alpha = \pm $ labels the two Fermi points, and the
dynamic structure factor is simply 
$ S^{RPA} ( {\bf{q}} , \omega ) = Z_{\bf{q}} \delta ( \omega - \omega_{\bf{q}})$,
with residue 
 $Z_{\bf{q}} = \frac{ | {\bf{q}} |}{2 \pi \sqrt{ 1 + F_{\bf{q}} } }$
 and collective mode
 $\omega_{\bf{q}} = 
 \sqrt{ 1 + F_{\bf{q}} } v_F | {\bf{q}} | $.
Here $F_{\bf{q}} = f_{\bf{q}} / ( \pi v_F )$ is the relevant dimensionless coupling.
(It is understood that in $d=1$ the vector ${\bf{q}}  $ has only one component.)
It is also easy to show\cite{Kopietzhab} that
in this case  $f^{RPA} ( {\bf{q}} , {\bf{v}}^{\alpha} \cdot {\bf{q}} ) = 0$, 
so that the first term in Eq.\ref{eq:S2} vanishes. Because of the $\delta$-function
dynamic structure factor,
the $\omega$-integration in  Eq.\ref{eq:S2} is trivial.
To perform the remaining (one-dimensional)  ${\bf{q}}$-integral,
we adopt the usual procedure\cite{Dzyaloshinskii74}
of replacing $f_{\bf{q}} \rightarrow f_0$ in the integrand  
and multiplying the integrand by an ultraviolet cutoff $e^{- | {\bf{q}} | / q_c }$.
The ${\bf{q}}$-integration is then elementary and we obtain
after straightforward algebra
 \begin{equation}
  Q^{\alpha}_1 ( t^{\prime \prime} )   =
  - (1 + \frac{\gamma}{2} )  \ln \left[ 1 + i ( \tilde{v}_F - v_F) q_c t^{\prime \prime}  
  \right]
  - \frac{\gamma}{2}  \ln \left[ 1 + i ( \tilde{v}_F + v_F )q_c  t^{\prime \prime}  
  \right]
  \; \; \; ,
  \label{eq:QTLMres}
  \end{equation}
where
  \begin{equation}
  \gamma  = \frac{ F_0^{2}}{2 \sqrt{ 1 + F_0} 
  \left( \sqrt{1 +  F_0} + 1 \right)^2 }
  \label{eq:gammapmdef}
  \end{equation}
is the  well-known anomalous dimension of the Tomonaga-Luttinger model,
and $\tilde{v}_F = \sqrt{1 + F_0} v_F$ is the
renormalized Fermi velocity.
For $\omega > 0$ the  eikonal result for the spectral function of the spinless
Tomonaga-Luttinger model can then be written as
 \begin{equation}
 {{A}}^{\alpha} ( {\bf{q}} , \omega )
  =  \frac{1}{\pi} Re \int_0^{\infty} d t^{\prime \prime} 
  \frac{e^{i 
  t^{\prime \prime}  
   [ 
 \omega - {\bf{v}}^{\alpha} \cdot {\bf{q}} + i 0^{+} ] 
 }}
  { \left[ 1 + i ( \tilde{v}_F -v_F ) q_c  t^{\prime \prime}  
  \right]^{1 + {\gamma}/{2} }
  \left[ 1 + i ( \tilde{v}_F + v_F ) q_c t^{\prime \prime}  
  \right]^{{\gamma}/{2} }}
  \label{eq:GretTLM}
  \; \; \; .
  \end{equation}
Note that the integrand vanishes as
$ ( t^{\prime \prime})^{ 1 + \gamma}$ for large $t^{\prime \prime}$.
Hence, for any finite interaction the integral in Eq.\ref{eq:GretTLM}
is convergent, and 
we may omit the convergence factor $i 0^{+}$ in the exponent.
Following Dzyaloshinksii and Larkin\cite{Dzyaloshinskii74}, we find 
that for small $| \omega - {\bf{v}}^{\alpha} \cdot {\bf{q}} | $
and $ 0 < \gamma < 1$ to leading order
$ {{A}}^{\alpha} ( {\bf{q}} , \omega )
 \sim
 A_0^{\alpha} + A_1^{\alpha} 
 | 
\omega - {\bf{v}}^{\alpha} \cdot {\bf{q}} |^{\gamma}$,
where $A_0^{\alpha}$ and $A^{\alpha}_1$ are finite complex numbers. 
This is in striking disagreement 
with the well-known bosonization result\cite{Dzyaloshinskii74}, which
predicts for $0 < \gamma < 1$ a singularity
of the type 
 ${{A}}^{\alpha} ( {\bf{q}} , \omega ) \propto
 | 
\omega - \sqrt{1 + F_0} {\bf{v}}^{\alpha} \cdot {\bf{q}} |^{\gamma - 1}$.
Thus, the real time eikonal method neither reproduces the
correct scaling behavior of the spectral function, nor does it
predict the correct location of its singularities.

\section{Conclusion}

In this work we have compared the real time eikonal method with
functional bosonization. 
In both methods the translationally invariant
Green's function $G$ of the many-body system is obtained 
by calculating first the Green's function ${\cal{G}}$ 
in a given background field, and then averaging the result
with respect to a suitably defined
Gaussian effective action, $G = \left< {\cal{G}} \right>$.
Although both methods produce very similar non-perturbative
expressions for ${\cal{G}}$, the final results
for the functinal average $ \left< {\cal{G}} \right>$ are certainly not equivalent.
In particular, in the case of the one-dimensional
Tomonaga-Luttinger model the real time eikonal method does not
even reproduce the correct scaling behavior of the Green's function.
We suspect that this shortcoming of the eikonal approach is 
related to the averaging in Minkowski time, which 
cannot be interpreted as averaging over an ordinary probability distribution. 
On the other hand, the imaginary time averaging 
performed in functional bosonization
is mathematically well defined,
and leads to an expression
for the Green's function which is exactly equivalent with
the result obtained via operator bosonization\cite{Houghton93,Castro94}. 

Because even at the level of the linearized theory the
real time eikonal expansion can lead to incorrect results,
we conclude that this method is not a good starting point
for studying the effects associated with the curvature of the Fermi surface 
on the low-energy behavior of the Green's function 
of non-Fermi liquids.
Inclusion of curvature effects is of particular importance in 
connection with the gauge field problem. 
It is therefore highly desirable to study the
effect of the non-linear terms in the energy dispersion 
on the leading (linearized) bosonization expression for the Green's function entirely 
within the framework of the imaginary time approach described
in Sec.\ref{sec:bos}. 
Work along these lines is in progress.

\section*{Acknowledgements}
\vspace{0.2cm}
I would like to thank K. Sch\"{o}nhammer for discussions and comments
on the manuscript. 
I have also profited from  discussions with G. E. Castilla, D. V. Khveshchenko, 
and V. Meden.

%

\begin{thebibliography}{99}
\bibitem{footnoteG}
We use the symbol ${\cal{G}}$ for the
Green's function in a given fixed external field, and write
$G = < {\cal{G}} >$ for the corresponding 
Green's function that is averaged with respect to an appropriate
effective action.
\end{thebibliography}
\end{document}